\documentclass[11pt]{article}

\usepackage{SIGACTNews}
\usepackage{amsmath}
\usepackage{url}

\newcommand{\sat}{\models}
\newcommand{\rimp}{\Rightarrow}

\renewcommand{\phi}{\varphi}

\title{SIGACT News Logic Column 18\\[2ex]
\textbf{Alternative Logics: A Book Review}\footnote{\copyright{} Riccardo Pucella, 2007.}} 
\author{Riccardo Pucella\\
Northeastern University\\
Boston, MA 02115 USA\\
riccardo@ccs.neu.edu}
\date{}

\begin{document}

\SIGACTmaketitle

A few years ago, Bill Gasarch, editor of the Book Review Column, sent
me a copy of the book \emph{Alternative Logics: Do Sciences Need Them?}
\cite{r:weingartner04}, edited by Paul Weingartner, and asked if I was
interested in reviewing it.  
The book is a collection of essays discussing whether there is a need
for logics other than classical logic in various areas of science.
Interestingly, I had just finished reading a copy of philosopher Susan
Haack's \emph{Deviant Logic} \cite{r:haack74}, concerning the
philosophical foundation of alternative logics.  
This was fortuitous, because Haack's monograph provided a wonderful
introduction to the essays collected in Weingartner's volume.

Let me start by giving a sense of the main question that underlies
both books.  Generally speaking, when one thinks of logic, one thinks
of classical first-order predicate logic. 
(Throughout, I will take propositional logic as a sublanguage of
first-order predicate logic.)
Mathematicians sometimes need to go up to second-order, that is,
allowing quantification over predicates, in order to express set
theory (at least, Zermelo-Fraenkel set theory) and thereby most of
modern mathematics. 
However, at various points in time, philosophers, mathematicians, and
scientists have advocated using logics different from classical logic,
arguing that the latter was not always appropriate. 
As a first example, take intuitionism. 
Intuitionism \cite{r:heyting71}, a philosophical position about the
meaning of mathematics, has profound implications on logic as a
framework to express mathematics. 
Without describing intuitionism in depth, one common feature of
intuitionism is that it is constructive. 
As a consequence,  the law of excluded middle is often
rejected in its full generality: intuitionism does not admit the
validity of $A\lor\neg A$; 
rather, $A\lor\neg A$ is true only when either $A$ can be established
explicitly, or $\neg A$ can be established explicitly. 
Fortunately, much of mathematics survives such a restriction in
proving power; but not all---some results are known only via a
nonconstructive proof. 
Thus, the question of which logic is the right logic for mathematics
impacts the mathematical results that can be proved.

Around the same time as intuitionism was proposed, another view of
classical logic was being questioned. 
Take propositional  logic.
There is a well known connection between propositional logic and
sets, an instance of Stone Duality \cite{r:stone36,r:johnstone82}. 
This connection views  propositional logic as
being the logic of \emph{events}, where an event is a set of states of
the world. 
An elementary proposition $A$ is identified with the set $A$ of states
where $A$ is true. 
A formula $A\land B$ is identified with the set of all states where
both $A$ and $B$ are true, which is equivalent to $A\cap B$, and
similarly $A\lor B$ is identified with the set of all states where
either $A$ or $B$ is true, which is equivalent to $A\cup B$. 
This correspondence implicitly depends on states behaving as
prescribed by Newtonian physics. 
In particular, Newtonian states satisfy a distributivity property
$A\cup(B\cap C) = (A\cup B)\cap (A\cup C)$
which is inherited by propositional logic. 
The advent of quantum mechanics has revealed that what we consider
states are not so well behaved, and therefore our identification of
propositional logic with the logic of events is somewhat misguided at the
quantum level. 
In particular, the distributive property does not always hold because
of superposition. 
Defining logics of events based on quantum states has led to many
proposals for quantum logics, starting with Birkhoff and von Neumann
\cite{r:birkhoff36}. 

These are but two instances of a general phenomenon. 
The situation has been compared to the situation in geometry
in the late 19th century. 
After two millennia where geometry was equated with Euclidean geometry,
the discovery that the parallel axiom was independent from the other
axioms led to the derivation of distinct geometries, as coherent as
Euclidean geometry. 
For a while, the only distinction of Euclidean geometry was that it
seemed to be the geometry of the real world. 
(Until, of course, Einstein blew a hole in this preconception.) 
Felix Klein's Erlangen programme was a consequence of this new view of
geometry, and was an attempt at understanding the plurality of
geometry.

The two examples I gave above illustrate that a similar discourse is
occurring about the status of logic. 
And this is the discourse reported in both Haack's and Weingartner's
volumes. 
Before proceeding with the discussion of the volumes, I want to point
out that I have borrowed the term \emph{plurality} from Beall and
Restall's \emph{Logical Pluralism} \cite{r:beall06}, which bears on the
topic covered in this column. Unfortunately, I have not yet obtained a
copy of this book. 
The teaser is intriguing, however:
\begin{quote}
This is our manifesto on \emph{logical pluralism}. We argue that the
notion of logical consequence doesn't pin down one deductive
consequence relation, but rather, there are many of them. In
particular, we argue that broadly classical, intuitionistic and
relevant accounts of deductive logic are genuine logical consequence
relations. We should not search for \emph{One True Logic}, since there
are \emph{Many}.\footnote{Taken from the book's web page at
  \url{http://consequently.org/writing/logical_pluralism/}.}
\end{quote}

\section*{Susan Haack: \emph{Deviant Logic}}

Haack's monograph is a pleasure to read, and provides a reasonably
approachable introduction to the topic of alternative logics from a
philosophical perspective.
At the risk of simplifying her presentation to the point
of caricature, Haack is interested in how
really different are non-classical logics.
From the onset, Haack distinguishes between:
\begin{itemize}
\item \emph{Extended logics}, which can be understood as classical
logic, extended with features necessary for reasoning about a
particular phenomenon not handled directly by classical logic. 
Modal logics such as temporal logic (operators to reason about time)
or epistemic logic (operators to reason about knowledge) are usually
taken to be extended logics.
Up to differences in syntax and vocabulary, theorems of classical
logic remain theorems of an extended logic.

\item \emph{Deviant logics}, which can be understood as capturing alternate
forms of reasoning.
Thus, for instance, Lucasiewicz's three-valued logic or intuitionistic
logic embody different logical principles than classical logic.
Roughly, some theorems of classical logic cease to be theorems in a
deviant logic, again up to differences in syntax and vocabulary.

\end{itemize}
Note that this classification is not as clear cut as one might hope.
There is a lot of wiggle room, for example, in the notion of
differences in syntax and vocabulary. (See Felleisen
\cite{r:felleisen91} for a computer science perspective on this last
topic.)

It is tempting to view extended logics as minor adjustments to our
logical apparatus, while viewing deviant logics as representing
\emph{rivals} of classical logic as a foundation.
However, this picture does not fare so well under close scrutiny, and
Haack spends the first half of her book trying to tease out to what
extent we should equate deviance (which is a technical concept) with
rivalry (which is a psychological attitude with respect to the 
position of a logic in science), and whether there is any meaning to
the notion of rivalry.

The second part of her book illustrates the notion of deviance by
examining five canonical examples in detail.
Besides intuitionism and quantum mechanics, which I discussed above,
Haack also examines future contingents (statements about the future
need not be necessarily true or false in the present, so what truth
value do we give them?), vagueness (predicates in the real world, for
instance  color, rarely seem strictly true or false), and singular
terms (references need not always denote; what do we do with formulas
that refer to the ``present king of France''?).
The discussion is more often than not illuminating. 

It may be useful at this point to ponder to what extent the above
classification has anything interesting to say as to how logic is used
in computer science. 
As readers of this column well know, logics in computer science are
used in a variety of ways, and my use of the plural here is fully
conscious.
The article \emph{On the Unusual Effectiveness of Logic in Computer
Science} \cite{r:halpern01d} gives an accessible overview of
application domains that have especially benefited from a logical
approach.
 Here are some (by no means disjoint or exhaustive) categories we can
readily identify:
\begin{itemize}
\item Specification and verification: The problem of specifying
 and formally verifying properties of systems is central to much of
 computer science. 
 Specification are often expressed in a form of modal logic, and
 verified by model checking: the logic serves as a formal language for
 writing down specifications, and verification amounts to checking
 that a specification $A$ is true in a representation $M$ of the
 system, that is, $M\sat A$ \cite{r:clarke99,r:huth99}. 
 Other verification approaches rely on interactive theorem proving for
 extremely expressive logics, usually higher-order---Isabelle
 \cite{r:paulson94} is an example of such a framework.

\item Artificial intelligence: The intent here is to model various
forms of reasoning, and the resulting logics are strongly related to
philosophical logics \cite{r:ramsay88}.
Additionally, the community studies modal logics for reasoning under
uncertainty (for instance, probability, or Dempster-Shafer belief
functions) \cite{r:halpern03e}, as well as non-classical logics for
capturing common sense reasoning, such as default logic
\cite{r:reiter78}.

\item Descriptive complexity: The complexity of a language (i.e., a
set of strings) can be studied by looking at how strings in the
language are characterized using formulas; different logics 
give rise to different languages being expressible
\cite{r:ebbinghaus95,r:immerman98,r:libkin04}. 
This is essentially a finite form of classical model theory, in which
mathematical structures are studied by looking at whether they can be
characterized by various fragments of first-order predicate logic. 
(A recent survey in \emph{SIGACT News} reports on progress in descriptive
complexity \cite{r:immerman05}.)

\item Computation: there has been many advances in understanding
  computations using intuitionistic logic.  The Curry-Howard
  isomorphism 
  tells us that proofs in propositional intuitionistic logic can be
  viewed as programs in a simply-typed lambda calculus that type check
  at the type expressed by the proposition \cite{r:girard88}. 
This correspondence can be pushed quite far, as witnessed by many
systems such as Coq \cite{r:coquand88} or Nuprl \cite{r:constable86}
that take an intuitionistic logic as foundation and can perform
\emph{program extraction} to automatically extract from the proof of a
property a program satisfying the property.\footnote{Interestingly,
classical logic can also be given a computational interpretation; the
law of excluded middle bears a strong relationship with non-local
control flow \cite{r:griffin90}.}
\end{itemize}

\newcommand{\brocc}{\ensuremath{\mathrel{\clubsuit}}}
\newcommand{\brocccons}{\ensuremath{\mathrel{\heartsuit}}}

Of course, just like any formal system, logic can be abused, and
computer science has seen a flourishing of logical systems whose
foundation can be questioned. 
Ramsey warns of such abuse in artificial intelligence; after arguing
that logical formalisms are necessary to provide a solid foundation
for artificial intelligence, he notes:
\begin{quote}
At the same time, it looks as though some of the papers using such
formalisms are merely disguising the poverty or unoriginality of the
work being reported. It seems as though you can make your program
respectable if you describe it using a dense logical notation, even
if it doesn't actually do anything interesting. \cite[p.vii]{r:ramsay88}
\end{quote}
Girard is less diplomatic in his somewhat iconoclastic appendix to
\emph{Locus Solum} \cite{r:girard01}: 
\begin{quote}
$\bullet$ \textsc{Broccoli logics}\\ 
Not as bad as paralogics, Broccoli logics are deductive. 
The basic idea is to find a logical operation or principle not yet
considered... which is not too difficult: call it \emph{Broccoli}. 
Then the Tarskian machinery works (here the symbol `\brocc'
stands for the syntactical Broccoli): 
\[\text{$A\brocc B$ is true if $A$ is true \textbf{Broccoli} $B$ is true.}\]
If you are smart enough to catch this delicate point, \textbf{Broccoli} is the meta of `\brocc'.
Broccoli is equipped with principles that have been never yet
considered, typically
\[(A \brocc B) \rimp (A \brocc (B \brocc B)) \]
and soundness and completeness are proved with respect to all
structures containing a constructor \brocccons{} enjoying 
\[ (a \brocccons b) \le (a \brocccons (b \brocccons
b)).\]
(Hint: to prove completeness, construct the \emph{free Broccolo}.) 
\cite[p.402]{r:girard01}
\end{quote}

(Girard, by the way, does present an alternative to classical logic in
\emph{Locus Solum}, where he introduces a new foundation for reasoning
based (very roughly) on a notion of games.  
This work builds on his previous work on linear logic \cite{r:girard87,r:girard88}, and
provides a third perspective on alternatives to classical logic. 
Linear logic is a form of substructural logic,  characterized
by not allowing arbitrary manipulations of formulas in premises of
deductions. 
For instance, linear logic does not allow one to duplicate a formula
in the premises of a deduction, so that each use of a formula in a
proof must be accounted for exactly. 
Substructural logics are not discussed in either Haack's book, or in
Weingartner's collection. Restall \cite{r:restall00} provides a  good
introduction to substructural logic. 
Note that the Lambek calculus which is at the basis of categorial
grammars in linguistics is a particularly weak form of substructural
logic \cite{r:lambek58,r:carpenter97}.)

Back to Girard's quote. One way to (constructively) read his criticism 
is that logic should bring something to the table. 
A logic is frequently prescribed syntactically, by giving formulas and
inferences rules and axioms that these formulas should satisfy. 
The semantics are often an afterthought, mathematical structures used
to interpret the truth of formulas validating the axioms.
The most useful logics, in computer science and elsewhere, tend to have
a semantics that is both intuitive (in that one can look at a model
and understand the significance of properties of that model) and
independently motivated. 
For instance, in logics for distributed computing and verification,
models are often simply derived from program executions, so that
models have a meaning independently from their use as structures in
which to interpret the logic.

\section*{Paul Weingartner: \emph{Alternative Logics}}

Weingartner's volume is an edition of essays on the topic of
alternative logics, with a particular focus on logics for science. 
The contributions are revisions of papers presented at the conference
``Alternative Logics: Do Sciences Need Them?'' of the Acad\'emie
Internationale de Philosophie des Sciences held at the Institut f\"ur
Wissenschaftstheory, Internationales Forschungszentrum Salzburg in May
1999. 
Shapere makes the point clearly in his contribution:
\begin{quote}
The question, ``Does science need a new logic?'' can be interpreted in
at least two ways. On the one hand, it can be understood as a question
of what Carnap would have called the `object-level': Do any
\emph{specific} areas of science today require a new logic in order to
solve \emph{specific} problems arising in those areas? [...] On the
other hand, the question can have a `metalevel' focus, namely, `Does
science (in the \emph{general} sense) require a new logic, or at least
a more persistent and competent application of the logic we have, in
order to understand its \emph{general} character and procedures?'(p.43).
\end{quote}

By and large, contributions in this collection cover both 
points given by Shapere. 
The collection has three parts, but the division is not a very crisp one. 
The first part covers the general concept of
alternative logics, essentially at the level of the first part of
Haack's monograph, with contributions discussing philosophical
implications of alternative logics. 
The second part focuses on discussions of alternative logics useful
for science as a whole, while the third part focuses on discussions
of alternative logics prompted by specific applications, such as
computer science or quantum mechanics.

Here is an outline of the contributions.

\begin{itemize}
\item[I.] \textbf{GENERAL TOPICS}

\item[1.] \textbf{Why is it Logical to Admit Several Logics}\\ Agazzi
  argues for there being many logics, in analogy with Klein's
  Erlangen's programme in geometry.
  More precisely, he argues that there is a sense in which there is a
  single logic, and a sense in which there is a plurality of logics,
and that the two sense can co-exist.

\item[2.] \textbf{Does Metaphysics Need a Non-Classical Logic?}\\
Quesada examines the relationship between metaphysics and logic. To
illustrate this relationship, note that one role of metaphysics is to
explicate what actually exists. This is reflected logically by the
extent of the existential and universal quantifiers. In this way,
metaphysical claims impact the interpretation of logical
constants. Quesada illustrates the relationship through the
metaphysics of Plotinus, Hegel, the empiricists (e.g., Mill), and
Routly (a proponent of noneism, advocating that the universe contains
nonexistent objects). 

\item[3.] \textbf{Logic and the Philosophical Interpretation of
Science}\\
Shapere revisits the Vienna circle's ``logistic'' programme of the first half
of the twentieth century, which was an attempt to understand science
(the whole of science, the enterprise of science) using the then
recent formal logic developed by Frege, Whitehead, and Russell, and
argues the reasons for the programme's failure. 

\item[4.] \textbf{How Set Theory Impinges on Logic}\\
Moster\'{\i}n examines how
logic and set theory are inextricably intertwined. 
In particular, while set theory (say, ZFC) is in flux, with many
unresolved questions, logic is thought to be independent and
essentially understood. However, as Moster\'{\i}n argues, second-order
logic (and up to a point, first-order logic) is as open as set
theory. In particular, every axiom of ZFC is expressible in
second-order logic, and with some work, every axiom of ZFC can be made
to correspond to a closed pure second-order logic formula, with the
property that the formula is valid if and only if the axiom
holds. Thus, the status of validities in second-order logic is
intrinsically linked to axioms of ZFC. More generally, what
second-order logic \emph{is} depends on what set theory is.

\item[5.] \textbf{Geometries and Arithmetics}\\
Priest re-examines the three a
priori sciences of Kant: arithmetic, geometry, and logic. 
It is by now accepted that geometry is not a priori, in the sense that
there is a plurality of geometries, and which one applies to our
external world is a contingent fact. 
Priest argues that arithmetic is similarly contingent, in that it is
conceivable to develop different models of arithmetic, alternatives to
the standard model---in particular, these models may be inconsistent,
but still useful. 
Interestingly, to reason about such models requires a paraconsistent
logic,\footnote{Very roughly speaking, a logic is paraconsistent if it
can be the underlying logic of an inconsistent (admitting both $A$
and $\neg A$ for some $A$) but nontrivial (not admitting all
formulas) theory.}
indicating that not only is arithmetic not a priori, neither is
logic. 

\item[6.] \textbf{Remarks on the Criteria of Truth and Models in
Science}\\ 
Del Re discusses the role of logic in the sciences of
Nature from the side of scientists. Del Re's view is that logic can be
taken to be the study of mental operations involved in the attainment
of a truth judgment, in particular, procedures by which knowledge is
obtained from premises and data, and criteria and rules for deciding
the validity of the knowledge so obtained. 
He further argues that more attention should be given to the role of
analogies and true versus reasonable statements. 

\item[7.] \textbf{Significant? Not significant? The Dilemma of Statistical
Inference}\\
Scardoni very briefly discusses  the question of
statistical inference and its role in science.

\item[II.] \textbf{ALTERNATIVE PROPOSALS}

\item[8.] \textbf{Outline of a Paraconsistent Category Theory}\\ Da
Costa, Bueno, and Volkov explore the definition of category theory
independently from set theory.
Starting from classical first-order predicate logic with
equality, they derive category theory as a theory in the logic. 
Then they show how to devise a paraconsistent category theory by
taking the underlying logic to be itself paraconsistent. 
A philosophical discussion of paraconsistency is included, necessary
to understand how the theory is to be used.

\item[9.] \textbf{Combinatory Logic, Language, and Cognitive
Representation}\\
Descl\'es proposes Church's combinatory logic as a foundation
for defining, analyzing, and comparing classical and non-classical
logical systems---a prelogic of sorts. 
The article is in fact a nice presentation of combinatory logic,
from a point of view different than the standard computer scientist
introduction to combinatory logic. 

\item[10.] \textbf{Extending the Realm of Logic: The Adaptive-Logic
Programme}\\
Batens proposes a logical approach for capturing actual reasoning,
which according to him requires both external dynamics (the possibly
of revising premises based on knowledge gained from the external
world) as well as internal dynamics (the possibility of revising
premises based on knowledge gained from the process of reasoning
itself). Abduction, the supposition of a premise based on the
explanatory power of that premise, is a typical example of internal
dynamics. 

\item[11.] \textbf{Comments on Jaako Hintikka's Post-Tarskian Truth}\\
Heinzmann
discusses Hintikka's IF-logic (for independence friendly
logic). IF-logic, very roughly, allows partially ordered quantifiers,
such as:
\[ \begin{aligned}[c]
    & \forall x\exists y\\
    & \forall z\exists u
   \end{aligned} ~S(x,y,z,u)
\]
which is an attempt at capturing  that the $y$ does not depend on the $z$, and
that the $u$ does not depend on the $x$. 
Heinzmann addresses Poincar\'e's foundational remarks about logic as
they pertain to IF-logic.

\item[III.] \textbf{ALTERNATIVE LOGICS MOTIVATED BY PROBLEMS OF
APPLICATION TO SCIENCE}

\item[12.] \textbf{Applied Logics for Computer Science}\\
Gochet and Gribomont describe two applications of first-order logic in
computer science.  
The first is the extension of first-order logic to reason about
program first advocated by Hoare, with assertions of the form
$\{A\}S\{B\}$, where $A$ and $B$ are formulas and $S$ is a
program. 
The meaning of such assertions is that executing program $S$ in a
state satisfying $A$ (the precondition) must either not terminate or
terminate in a state satisfying $B$ (the postcondition). 
These assertions come with inference rules that allow one to derive
assertions from assertions about subprograms. 
The inference rule for looping is especially interesting:
\[ \frac{\{I\land
B\}S\{I\}}{\{I\}\mathsf{while}~B~\mathsf{do}~S\{I\land\neg B\}}\]
which uses an invariant $I$ preserved by every iteration of the loop.
The problem of constructing such invariants during a proof of an
assertion is examined by Gochet and Gribomont.
The second application of first-order logic they describe is logic
programming, via a rather nice tutorial illustrating the core ideas of
this programming paradigm.

\item[13.] \textbf{Stochastic \emph{versus} Deterministic Features in
Learning Models}\\
Stamatescu gives an overview of the debate on the role of randomness
and stochastic phenomena in scientific inquiry.
Roughly speaking, the two sides of the debate take the view
that randomness should either be treated as  ``statistical noise'', or
be taken into account directly as a process at work in the model. 
This debate is illustrated in the context of learning theory.

\item[14.] \textbf{Praxic Logics}\\
Finkelstein and Baugh contribute the first article on quantum logic in
this collection. 
They mainly argue for a variant semantics for quantum theory, based on
the observation that ``pre-quantum physics describes object but
quantum physics represents actions'' (p. 218). They propose a logic
for quantum states corresponding to this variant semantics. 

\item[15.] \textbf{Reasons From Science for Limiting Classical
Logic}\\
Weingartner, after reviewing some existing logical systems for
quantum mechanics, proposes an alternative quantum logic that can be
viewed as a restriction of first-order predicate logic. 
Very roughly speaking, inference is restricted so that propositional
variables in valid schemas of predicate logic cannot be replaced by
arbitrary formulas, but rather must obey a replacement criterion. 
Similarly, a valid predicate logic formula cannot in general be
``reduced'' to equivalent simpler formulas (e.g., $C\land C$ to $C$). 
Weingartner argues that such an approach provides a viable logic
system for quantum mechanics.

\item[16.] \textbf{The Language of Interpretation in Quantum Physics
and Its Logic}\\
It has been widely accepted since the days of the Copenhagen
interpretation of quantum mechanics that there is no good language for
describing what happens at the quantum level. Omn\'es argues that
there \emph{is} a convenient language for expressing interpretation.

\item[17.] \textbf{Why Objectivist Programs in Quantum Theory Do Not
Need an Alternative Logic}\\
Cordero critiques the logical turn in foundational quantum theory by
questioning some assumption of this programme. 
This questioning highlights the extent to which proposed quantum
logics still embody classical intuitions.
He then proceeds to show that these assumptions are mainly dropped
from three of the most developed objectivist approaches to quantum
theory. 

\item[18.] \textbf{Does Quantum Physics Require a New Logic?}\\
Mittelstaedt argues that there is no pluralism of logical systems
corresponding to different fields of experience (for instance,
classical reality versus quantum reality), and that instead there is a
hierarchy of logics, with at its base a ``true'' logic of propositions
about physical systems, which he takes to be a quantum logic.

\item[19.] \textbf{Experimental Approach to Quantum-Logical
Connectives}\\
Stachow devises a process-based semantics for Mittelstaedt's quantum
logic (see 18 above), starting from experiments for elementary
propositions, and developing experiments yielding logical connectives.

\item[20.] \textbf{From Semantics to Syntax: Quantum Logic of
Observables}\\
The original presentation of quantum logic by Birkhoff and von Neumann
is really an algebraic presentation of a quantum logic, in the same
way that Boolean algebras are an algebraic presentation of
propositional logic. 
Vasyukov attempts a syntactical reconstruction of quantum logic
corresponding to an algebraic semantics. 

\item[21.] \textbf{An Unsharp Quantum Logic from Quantum
Computation}\\
Cattaneo, Dalla Chiara, and Giuntini take a first step in  deriving a
quantum logic with a semantics informed by quantum computation. 
Roughly speaking, the quantum equivalent of logic gates (operating on
qbits, the primitive elements of quantum computation, rather than bits)
are taken to provide semantics for the logical connectives. 
The resulting logic seems to be a weak form of quantum logic.

\item[22.] \textbf{Quantum Logic and Quantum Probability}\\
Beltrametti first reviews algebraic models of events in classical and
quantum logic, as well as variations in properties of
probability in classical and quantum systems. He then proposes a
common extension of classical probability theory and quantum
probability theory, by taking convex sets of states as building
blocks. 

\item[23.] \textbf{Operator Algebras and Quantum Logic}\\
R\'edei examines the process of deriving a logic from an algebraic
semantics, in particular when the algebra of events is taken to be a
general class of non-Boolean lattices arising naturally from certain
quantum systems.

\end{itemize}

As a whole, the papers in the collection tend to be short, and not
quite self-contained---the mathematics is often kept short, statements
are not proved, and some literature chasing is perforce
necessary to understand a paper fully.
In particular, had I not gone through Haack's monograph before reading
the collection, many of the subtleties would have over my head. 
There is enormous variation as to the level of technical details
present in each contribution, from the more historical and
philosophical pieces to the more mathematically-oriented ones.
By and large, the more mathematically challenging pieces are those
dealing with set theory and with quantum logics.
This probably fits correctly with the intended audience, philosophers
of science.
It is not clear to what extent this collection will speak to an even
theoretically-minded computer scientist. 

Logic has been called ``the calculus of computer science''.
Weingartner's volume is not calculus for engineers, but calculus for
mathematicians. 
It does not directly impact the daily life of practitioners, but may
contribute to a greater understanding of the foundations.

\end{document}